\documentclass{article}
\usepackage[utf8]{inputenc}
\usepackage{geometry}
\usepackage{tabularx}
\usepackage{booktabs}
\usepackage{graphicx}
\usepackage{hyperref}
\usepackage{float}
\usepackage[
backend=biber,
style=alphabetic,
sorting=ynt
]{biblatex}

\date{}
\title{A Systematic Literature Review on Cyber Threat Hunting}
\author{Zichen Wang (1254236) University of Guelph, Ontario}

\begin{document}

\maketitle

\section*{Abstract}
Since the term "Cyber threat hunting" was introduced in 2016, there have been a rising trend of proactive defensive measure to create more cyber security. This research will look into peer reviewed literature on the subject of cyber threat hunting. Our study shows an increase in the field with methods of machine learning.\\
Keywords: Cyber threat, Cyber security, threat hunting , security system, data driven, Intel, analytic driven, TTPs \\
\newpage
\tableofcontents
\newpage
\section{Introduction}
The internet has become a fundamental part of the modern society, tremendous amount of valuable data and money are transferred through the internet, and cyber security is even more important today than ever. To combat cyber threats, one of the emerging new methodology is cyber threat hunting. Cyber Threat hunting is a proactive cyber defence activity. Instead of waiting for the cyber threat to compromise the system, the cyber security personnel will proactively hunt for the potential cyber threat. Although cyber threat hunting is practiced in many companies, these companies are not yet satisfied with the effectiveness of cyber threat hunting.\\
For cyber threat hunting, the most commonly known categories are: First, the data-driven hunting, which is triggered by data observation [1]\cite{1}; Second, the Intel-driven hunting, identifies threat intelligence information [2]\cite{2}; Third, the techniques, tactics, and procedures, (TTP) driven hunting, searching for threat tactics, techniques and procedures.[3] \cite{3}\\
Cyber threat huntingis used as baseline for next generation security operation center. [4]\cite{4} Cyber threat hunting has also been used in industrial control systems. [5] \cite{5}[6]\cite{6}

\subsection{Prior Research}
In recent papers, we can see some new methods used for cyber threat hunting. There is a study on evidence based classification method for cyber threat hunting by Matthew Beechey team. [7]\cite{7}\\
In the study of another team, Fengyu Yang team [8]\cite{8}, cyber threat hunting is more flexible becuase their study is somewhat a hybrid cyber threat hunting.\\
The study of Abbas Yazdinejad team in 2022 [9]\cite{9}, where a deep learning model is used for cyber threat hunting in industrial internet. As we have seen in the paragraph above, industrial internet is a major part of the cyber threat hunting.\\
In the 2021 paper by Renzheng Wei team, we see them using the method of neural network for cyber threat hunting.[10]\cite{10}\\
Similarly, in the 2022 paper by Chung-Kuan Chen team, they inplement machine learning as a way for cyber threat hunting.[11]\cite{11}\\
\subsection{Research Goals}
The purpose of this research is to study existing studies of cyber threat hunting, and our goal is to anaylyze these studies and find the modern trend for cyber threat hunting. To better illustrate this research, see Table 1.
\subsection{Contribution and Layout}
\begin{itemize}
\item This research is complementary to existing cyber hunting researches, and is meant to provide some insights for other researchers looking into literature study of cyber threat hunting.\\
\item We identify 95 papers related to the study of cyber threat hunting, with the latest in year 2022.\\
\item Then from these papers, we selected 25 essential papers for further in depth study, because they meet with the keywords we set for these papers. And thus these papers can provide essential; information in our comparative study.\\
\item We will conduct a study of the data in these 25 papers.\\
\item And we will provide a meta-analysis of the methods introduced in these papers.\\
\end{itemize}
In this research, we make presentations and provide a possible trend for future cyber hunting studies.\\

\begin{table}[h]
\centering
    \caption{Research Questions
    }
    \label{crouch}
    \begin{tabular}{  l  p{3.4cm}  p{3.4cm} }
        \toprule
\textbf{Research Questions}      
& \textbf{Discussion}   
\\\midrule
What are the latest cyber threat hunting methods?
& In this research, we have observed data-driven, Intel-driven and TTP-driven.\\\hline
How is cyber threat hunting used to improve cyber security?    
& In this research, we believe cyber threat hunting can create security for industrial networks, and possibly provide a higher level of security to other fields.\\\hline
What new methods are available for cyber threat hunting?     
&  In this research, we have observed deep learning, machine learning and neural network used for cyber threat hunting.\\\hline
    \end{tabular}
\end{table}

\newpage
\section{Research Methodology}
To achieve answer the research questions, we conduct the SLR. In this section, we discuss the planning, conducting and reporting of the review in iterations to allow for a thorough understanding of this SLR.
\subsection{Selection of Primary Studies}
The selection process involves using different searching strings in the different database platforms. In our study, we used search string as "cyber threat", "threat hunting", "data-driven".\\
The platforms searched were:\\
- IEEE Xplore Digital Library\\
- ScienceDirect\\
- SpringerLink\\
- ACM Digital Library\\
- Google Scholar\\
The searches were run against the title of Cyber Threat Hunting, keywords or abstract, depending on different databases. All these searches were conducted on 15th October 2022, and we processed all recent studies in the databases that were published before this date. Then we filter all the papers we found in these databases through our inclusion/exclusion criterias.\\
\subsection{Inclusion and Exclusion Criteria}
Studies to be included in this SLR must report empirical methodology of cyber-threat hunting and could be papers on case studies, new technology and commentaries on the development of existing Cyber-threat hunting methods such as data-driven or Intel-driven. They must be peer-reviewed and written in English.

\begin{table}[h]
\centering
    \caption{Inclusion and Exclusion criteria
    }
    \label{crouch}
    \begin{tabular}{  l  p{3.4cm}  p{3.4cm} }
        \toprule
\textbf{Inclusion}      
& \textbf{Exclusion}   
\\\midrule
Paper shows empirical use of method.
& Paper focuses on cyber threat.\\\hline
Paper contains information of threat hunting.
& Data log of information\\\hline
Paper is written in English.
& Paper not in English.\\\hline
    \end{tabular}
\end{table}
\subsection{Selection Results}
There were a total of 937 studies identified from the initial keyword searches on the selected platforms. Then we excluded the ones that do not meet our need for this research. We get a total of 30 papers. After which we selected 25 from the 30 viable papers.\\
\subsection{Quality Assessment}
An assessment of the quality of primary studies was made according to the guidance.\\
\begin{table}[h]
\centering
    \caption{Excluded Papers
    }
    \label{crouch}
    \begin{tabular}{  l  p{3.4cm}  p{3.4cm} }
        \toprule
\textbf{Checklist for Stages}      
& \textbf{Excluded Studies}   
\\\midrule
Stage 1 Cyber Threat Hunting
& [26]\cite{26}\\\hline
Stage 2 Context
& [27]\cite{27} \\\hline
Stage 3 Method Application
& [28]\cite{28}\\\hline
Stage 4 Threat type
& [29]\cite{29}\\\hline
Stage 5 Data reliability
& [30]\cite{30}\\\hline
    \end{tabular}
\end{table}
\\
Stage 1: Cyber Threat hunting, the paper must be on the subject of threat hunting or emphasizing a method for threat hunting.
\\
Stage 2: Context, how the cyber threat hunting method is utilized in this paper.
\\
Stage 3: Method Application, if this method is one of the new methods or the traditional.
\\
Stage 4: Threat type, if this paper provides enough information about the cyber threat.
\\
Stage 5: Data reliability, if the data is reliable
\\
In this way, we can make sure this research is more effective and all the data that we extract from the papers are reliable enough for further analysis.\\
\subsection{Data Extraction}
All the papers in the reference have now passed the quality assessment, so now we extract all their study data from their papers, and we can divide the study data into these categories.\\
Context data: Information about the purpose of the study.\\
Qualitative data: Findings and conclusions provided by the authors.\\
Quantitative data: When applied to the study, data observed by experimentation and research.\\
\subsection{Data Analysis}
To meet the objective of answering the research questions, we
compiled the data held within the qualitative and quantitative data categories. And we conducted a meta-analysis of the papers.\\
\begin{figure}[h]
\centering
\includegraphics[width=12cm]{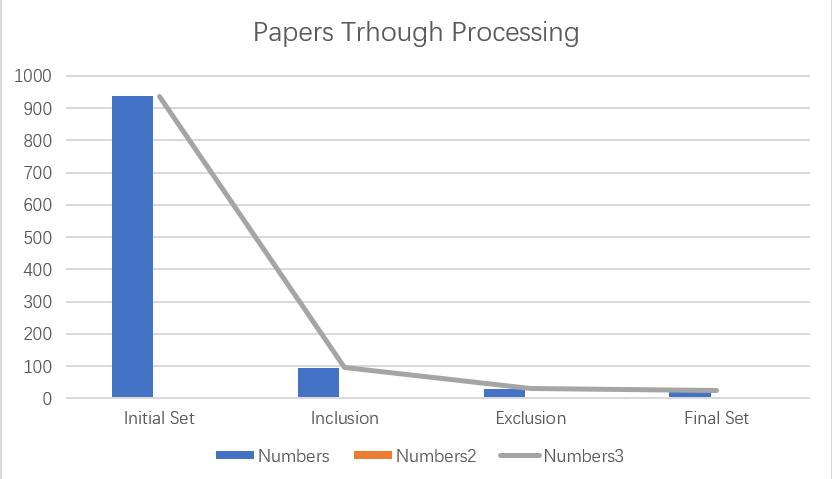}
\caption{Paper Through Processing}
\label{Figure}
\end{figure}
\newpage
\subsubsection{Publication over time}
The numbers of publication is increasing over the years. There was only 3 during 2018-2019. Another 5 during 2019-2020 and 7 during 2020-2021. And There were 10 between 2021-2022. \\
So we can envision there to be way more study in this area.\\
\begin{figure}[h]
\centering
\includegraphics[width=12cm]{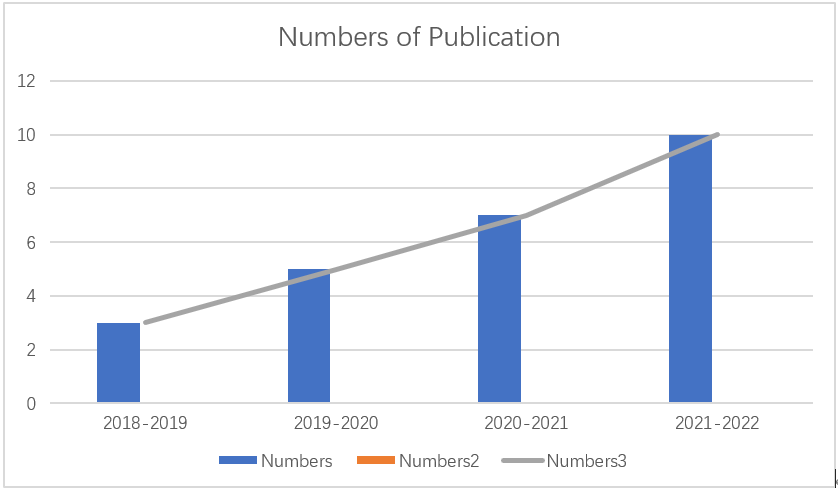}
\caption{Number of Publications}
\label{Figure}
\end{figure}
\newpage
\subsubsection{Significant Keyword Count}
In order to summarize all these papers, we decide to conduct a keyword search to find the most common themes among these papers. And obviously, Cyber threat and threat hunting were the top of the table.\\
\begin{table}[h]
\centering
    \caption{Keywords Count
    }
    \label{crouch}
    \begin{tabular}{  l  p{3.4cm}  p{3.4cm} }
        \toprule
\textbf{Keywords}      
& \textbf{Count}   
\\\midrule
Cyber Threat
& 973\\\hline
Threat Hunting
& 550\\\hline
Deep Learning
& 151\\\hline
Intelligence
& 125\\\hline
security system
& 124\\\hline
    \end{tabular}
\end{table}
\newpage

\section{Findings}
Each primary research paper was read in full and relevant qualitative and quantitative data was extracted and summarized in [Table 5-13]. All primary studies focuses on the application and different methods of cyber threat hunting. 
The focus of all papers is recorded below in [Table 5-13].\\
Each paper's theme and focus was further grouped into five primary categories, Machine Learning, TTP, Data, Intel and Neural Network.\\
Studies that had a focus concerning deep learning, grouped together into the machine learning category.\\
Figure 3 shows the percentages of different themes of the 25 primary studies, all of these have passed the quality assessment to be included in the data analysis.
The Most common themes identified in the primary studies are the studies of Machine Learning at 29 percent in the application of Cyber Threat Hunting, and in most of these cases, the use of Deep Learning is the focus.
Tied in the first place,the other most common theme of the studies is TTP, the analyses of tactic , technique and procedure at 29 percent.
Data is the third commonest theme at 17 percent, these are the most traditional studies of the cyber threat hunting. And one of these studies is the hybrid study of combining data-driven with other focus different studies.
Neural Network and Intel-driven are tied as the fourth most common focus in these papers at 13 percent. In a sense it is similar to Machine Learning but the focus is different.\\
\newpage
\begin{figure}[!htbp]
\centering
\includegraphics[width=16cm]{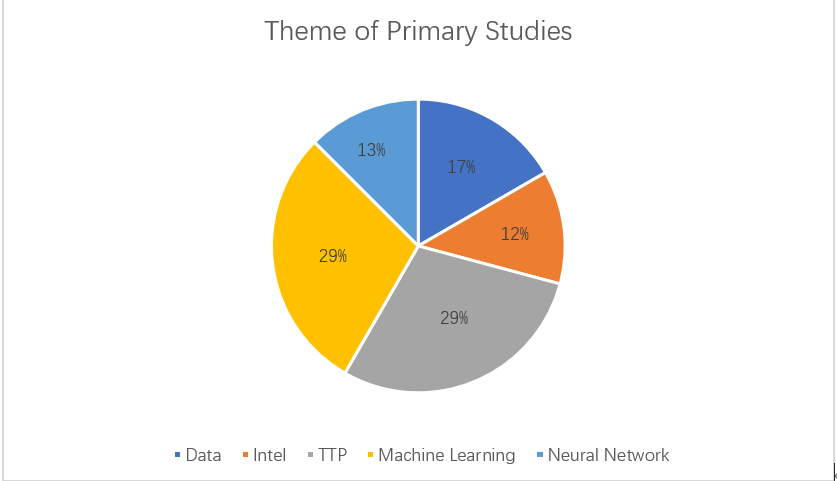}
\caption{Paper Through Processing}
\label{Figure}
\end{figure}
\newpage
\begin{table}[!htbp]
\centering
    \caption{Main findings and primary theme of the papers
    }
    \label{crouch}
    \begin{tabular}{  l  p{3.4cm}  p{3.4cm} }
        \toprule
\textbf{Primary Study}      & \textbf{Key Qualitative and Quantitative Data Reported}  & \textbf{Type of Cyber Threat Hunting} 
\\\midrule
1
& 
This paper presents a new automated threat assessment system that relies on the analysis of continuous incoming feeds of Sysmon logs to classify software in different threat levels and augment cyber defensive capabilities through situational awareness, prediction, and automated courses of action.
& Data\\\hline
2
& This paper explores query synthesis mechanism that automatically synthesizes a TBQL query for hunting, a system that facilitates threat hunting in computer systems using OSC
& Intel\\\hline
3
& In this article, the author challenge this model using an attack campaign mimicking APT29, a real-world threat, in a scenario designed by the MITRE Corporation. 
& TTP\\\hline
\end{tabular}
\end{table}
\newpage

\begin{table}[!htbp]
\centering
    \caption{Main findings and primary theme of the papers
    }
    \label{crouch}
    \begin{tabular}{  l  p{3.4cm}  p{3.4cm} }
        \toprule
\textbf{Primary Study}      & \textbf{Key Qualitative and Quantitative Data Reported}  & \textbf{Type of Cyber Threat Hunting} 
\\\midrule
4
& In this paper the author covered the most effective ways to manage a multi-million security infrastructure, using intelligent SIEM solutions and Threat Hunting techniques to increase the SOC efficiency and decrease the overall security cost.
& Intel\\\hline
5
&ICS-THF consists of three stages, threat hunting triggers, threat hunting, and cyber threat intelligence. The hunting stage uses a combination of the MITRE ATTCK Matrix and a Diamond model of intrusion analysis to generate a hunting hypothesis and to predict the future behaviour of the adversary.
& TTP \\\hline
6
& In this paper,the author take a novel machine learning algorithm which combines fuzzy logic with Bayesian inference to produce an optimized fuzzy model for identifying threats in cybersecurity data-sets
& Machine Learning \\\hline
\end{tabular}
\end{table}
\newpage

\begin{table}[!htbp]
\centering
    \caption{Main findings and primary theme of the papers
    }
    \label{crouch}
    \begin{tabular}{  l  p{3.4cm}  p{3.4cm} }
        \toprule
\textbf{Primary Study}      & \textbf{Key Qualitative and Quantitative Data Reported}  & \textbf{Type of Cyber Threat Hunting} 
\\\midrule
7\cite{7}
& 
The proposed approach is evaluated on a recent, challenging scenario of network security attacks and compared against multiple feature selection techniques. Based on the selected features, cyber threat classification analysis is performed using seven state-of-the-art ML classification algorithms
& 
Data
\\\hline
8\cite{8}
& 
In this paper, evaluations based on the adversarial engagement designed by DARPA prove that the platform can effectively hunt sophisticated threats, quickly restore the attack path or assess the impact of attack.
& 
Hybrid\\\hline
9\cite{9}
& 
In this paper, an ensemble deep learning model that uses the benefits of the Long Short-Term Memory (LSTM) and the Auto-Encoder (AE) architecture to identify out-of-norm activities for cyber threat hunting in IIoT is proposed.
& 
Machine Learning\\\hline
    \end{tabular}
\end{table}
\newpage
\begin{table}[!htbp]
\centering
    \caption{Main findings and primary theme of the papers
    }
    \label{crouch}
    \begin{tabular}{  l  p{3.4cm}  p{3.4cm} }
        \toprule
\textbf{Primary Study}      & \textbf{Key Qualitative and Quantitative Data Reported}  & \textbf{Type of Cyber Threat Hunting} 
\\\midrule
10
& 
The author design a graph neural network architecture with two novel networks: attribute embedding networks that could incorporate Indicators of Compromise (IOCs) information, and graph embedding networks that could capture the relationships between IOCs.
& 
Neural Network\\\hline
11
& 
In this study, the author share their past experiences in building machine learning-based threat-hunting models. It is a machine learning-based anomaly detection and threat hunting system which leveragesnatural language processing (NLP) and graph algorithms.
& 
Machine Learning\\\hline
12
& 
In this research,the author investigated that threat hunting in conjunction with cyber deception and kill chain has countervailing effects on detecting SCADA threats and mitigating them
& TTP

\\\hline
    \end{tabular}
\end{table}
\newpage
\begin{table}[!htbp]
\centering
    \caption{Main findings and primary theme of the papers
    }
    \label{crouch}
    \begin{tabular}{  l  p{3.4cm}  p{3.4cm} }
        \toprule
\textbf{Primary Study}      & \textbf{Key Qualitative and Quantitative Data Reported}  & \textbf{Type of Cyber Threat Hunting} 
\\\midrule
13\cite{13}
& 
This paper applies a fundamentally different approach to the problem, exploiting Isolation Forests, an unsupervised machine learning algorithm in a new context. One of the most important advantages of the algorithm is that it can identify and record novel intrusion models.
& 
Machine Learning
\\\hline
14\cite{14}
& 
WILEE uses the implementations along with other logic, also written in the DSL, to automatically generate queries to confirm (or refute) any hypotheses tied to the potential adversarial workflows represented at various layers of abstraction.
& 
TTP
\\\hline
15\cite{15}
& 
The new cyber reasoning methodology introduces an operational semantics that operates over three sub spaces -- knowledge, hypothesis, and action -- to enable human-machine co-creation of threat hypotheses and protective recommendations
& 
Intel
\\\hline
    \end{tabular}
\end{table}
\newpage
\begin{table}[!htbp]
\centering
    \caption{Main findings and primary theme of the papers
    }
    \label{crouch}
    \begin{tabular}{  l  p{3.4cm}  p{3.4cm} }
        \toprule
\textbf{Primary Study}      & \textbf{Key Qualitative and Quantitative Data Reported}  & \textbf{Type of Cyber Threat Hunting} 
\\\midrule
16\cite{16}
& 
In this paper,effective CTI technique is required to obtain knowledge from external data sources and
combine it with internal sources to enhance the hunting capabilities. Then, using the optimal data analysis
technique is needed for the CTH approach to obtain valuable insights into abnormal patterns in running
activities in the early stages.
& 
Data
\\\hline
17\cite{17}
& 
The author identify attack patterns, tactics, and techniques that exploit these CVEs and also uncover a disparity in how much linked information exists for each of these CVEs. This prompts us to further inventory BRON's collection of sources to provide a view of the extent and range of the coverage and blind spots of public data sources.
& 
TTP
\\\hline
18\cite{18}
& 
In this paper, the author propose a detailed analysis support system for threat hunting using three key ideas: making TTP icons to help translate events, similarity value visualization, relevance visualization between log entries.
& 
TTP
\\\hline
    \end{tabular}
\end{table}
\newpage
\begin{table}[!htbp]
\centering
    \caption{Main findings and primary theme of the papers
    }
    \label{crouch}
    \begin{tabular}{  l  p{3.4cm}  p{3.4cm} }
        \toprule
\textbf{Primary Study}      & \textbf{Key Qualitative and Quantitative Data Reported}  & \textbf{Type of Cyber Threat Hunting} 
\\\midrule
19\cite{19}
& 
The solution in this study is a central and open-source implemented using different opensource technologies, e.g., Elasticsearch, Conpot, Metasploit, Web Single Page Application (SPA), and a machine learning analyser.
& 
Machine Learning
\\\hline
20\cite{20}
& 
A system that facilitates cyber threat hunting in computer systems using open-source Cyber Threat Intelligence (OSCTI)
& 
Intel
\\\hline
21\cite{21}
& 
In this paper, we proposed a deep learning-based model for Ethereum threat hunting. The model applies a deep neural network for attack detection and uses a combination of machine learning algorithms (unsupervised with supervised algorithms) for attack classification.
& 
Neural Network
\\\hline
    \end{tabular}
\end{table}
\newpage
\begin{table}[!htbp]
\centering
    \caption{Main findings and primary theme of the papers
    }
    \label{crouch}
    \begin{tabular}{  l  p{3.4cm}  p{3.4cm} }
        \toprule
\textbf{Primary Study}      & \textbf{Key Qualitative and Quantitative Data Reported}  & \textbf{Type of Cyber Threat Hunting} 
\\\midrule
22\cite{22}
& 
In this paper, data discovered was used in an experiment that applied and evaluated the effectiveness of machine learning and decision-making algorithms in the method proposed to prioritize hypotheses in the screening phase
& 
Machine Learning
\\\hline
23\cite{23}
& 
This study describes Threat Hunting in an ecosystem as the constructive, analyst-driven scanning mechanism for attackers TTP. The model has been checked for real-world data sets using a variety of threats.
& 
TTP\\\hline
24\cite{24}
& 
This paper explores a novel flow analyzer based on DL architecture named LSTMSCAE-AGRU is designed by combining Long Short-Term Memory Stacked Contractive AutoEncoder (LSTMSCAE) with Attention-based Gated Recurrent Unit (AGRU) at the control plane.
& 
Machine Learning\\\hline
\end{tabular}
\end{table}
\newpage
\begin{table}[!htbp]
\centering
    \caption{Main findings and primary theme of the papers
    }
    \label{crouch}
    \begin{tabular}{  l  p{3.4cm}  p{3.4cm} }
        \toprule
\textbf{Primary Study}      & \textbf{Key Qualitative and Quantitative Data Reported}  & \textbf{Type of Cyber Threat Hunting} 
\\\midrule
25\cite{25}
& 
We demonstrate the method on large-scale real-world data, where it outperforms the unsupervised approach (Isolation Forest and Lightweight Online Detector of Anomalies), the supervised approach (Random Forest) and the traditional similarity search algorithm (kNN).
& 
Data
\\\hline
    \end{tabular}
\end{table}
\newpage

\section{Discussion}
The initial research into Cyber Threat Hunting yields more insights into the general consensus on the subject. A deeper look into the studies show that one of the major issues many cases of cyber threat hunting has shown is the lack of efficiency as such a hunting process is time consuming and required human effort.\\
However, with recent developments in artificial intelligence and potentially more open data bases on these subjects, machine learning, deep learning and neural networking is becoming one of the most common theme in these recent papers. As we can see in Figure 2 and Table 5-13, the number of papers related to cyber threat hunting and machine learning is slowly but surely increasing.\\
The researchers of these papers used established open source platforms, such as the  open-source Cyber Threat Intelligence (OSCTI) such as Case [2]\cite{2} and Case [20]\cite{20}. There are several reasons for researcher to use such platforms. The OSCTI is about sharing intel on cyber threats, therefore for the studies that focuses on intel-driven study, these open source platform is most useful. Machine learning can also use open source platforms as we can see in Case [19]\cite{19}.\\
Deep learning and Machine Learning are used to create models for threat hunting, as we can see in Case [6]\cite{6}, Case [9]\cite{9},Case [11]\cite{11}. The other papers build a machine learning algorithm for decision making, in order to maximize the efficiency of cyber threat hunting, as we can see in Case [13]\cite{13}, Case [22]\cite{22}, and Case [21]\cite{21}, which also leaned towards neural networking.\\
\subsection{RQ1: What are the latest cyber threat hunting methods?}
The latest cyber threat hunting methods are the uses of Machine Learning, Deep Learning and Neural Networks. As we can see in Figure 3, and we can see a lot more about the latest cyber threat hunting methods from Table 5-13.\\
Most recent methods includes Neural Networking such as Case [10]\cite{10} and Case[21]\cite{21}. \\
The far more popular new method is machine learning.\\
\subsection{RQ2: How is cyber threat hunting used to improve cyber security?}
In many ways, TTP is the best to explain why cyber threat hunting is used to improve cyber security.\\
1, Blind spots in public data sources, as seen in Case[17].\\
2, Hypothesis of future cyber threat, as stated in Case[5], Case[12]\cite{12}, with TTP-driven cyber threat hunting, the security can predict and prepare for potential threat before the system is under attck.\\
\subsection{RQ3: What new methods are available for cyber threat hunting? }
According to the primary studies, we can see a rising trend of using hybrid types of cyber threat hunting, by combining machine learning with TTP, the system will be more capable of predicting how the cyber threat will attack.\\
The latest studies suggested that the most commonly accepted applications for machine learning/neural networking were as follow:\\
1, Machine Learning algorithm for Decision Making in the neural networking. As seen in Case [6]\cite{6} , Case [11]\cite{11} , Case[13]\cite{13} and Case [21]\cite{21} and [22]\cite{22}.\\
2, Machine Learning analyzer, AS in Case [19]\cite{19}.\\
3, Threat hunting model based on Machine Learning. Commonly seen in machine learning cases such as Case [6]\cite{6}, Case [9]\cite{9},Case [11]\cite{11}
\newpage

\section{Future Research Directions of Cyber Threat Hunting
}
Based on the results of this literature analysis and our observations, we present the following research directions that are worth further investigations.
\subsection*{Deep Learning and TTP}
The hybrid study of TTP and Deep Learning, where the system will learn from Open Sources Cyber Threats and project the future TTP for cyber threat hunting.
\subsection*{Neural Network}
The neural network, which is linked commonly to the crafting of a machine learning algorithm in order to create the clear decision making pathway for the neural networking system. Further study could prove more efficient bsed on past ML models as seen in Case [21], Case [22].
\subsection*{Industrial ML}
Machine Learning combined with industrial cyber security. This is a potential projection because most commonly known industrial cyber threat hunting are still based on classifier which s data-driven or TTP-driven as in Case [5]. By combining the machine learning methodology into classifier for industrial cyber threat hunting could greatly improve their efficiency.
\\
\newpage

\section{Conclusion and Future Work}
This research has identified available recent research on how Cyber Threat Hunting can be used to increase the level of security of all cyber networks. \\
The initial search identified 937 studies with the initial keyword search on all the selected platforms. Those keywords includes Cyber Threat, Threat Hunting and TTP.\\
This reseach also highlighted potential future directions for research in the area.
\subsection*{Potential Research Agenda 1}
The research concerning Deep Learning and TTP-driven Cyber threat hunting. With Deep Learning, TTP can be so much more efficient.
\subsection*{Potential Research Agenda 2}
The research into Neural Networking and creating models for Cyber Threat Hunting. It may be possible to create a fully automated system for Cyber Threat Hunting. Case[10]
\subsection*{Potential Research Agenda 3}
The research into Industrial Machine Learning. As we can see, industrial cyber security is the most common theme in TTP-driven hunts. Case [5] 
\newpage

\section*{Declaration and Interest}
None
\section*{Acknowledgement}
None
\newpage

\section*{References}
\printbibliography[
heading=bibintoc,
title={Whole bibliography}
] 

\end{document}